\documentclass[aps,pra,twocolumn,showpacs,groupedaddress]{revtex4}

\usepackage{graphicx,bm}
\usepackage{amsmath,amssymb}
\usepackage{graphicx,amsfonts,amsbsy}

\renewcommand{\a}{\alpha}
\newcommand{\e}{\epsilon}
\newcommand{\f}{\phi}
\newcommand{\g}{\gamma}
\newcommand{\ff}{\varphi}

\newcommand{\m}{\mu}
\newcommand{\n}{\nu}
\newcommand{\p}{\pi}
\newcommand{\q}{\theta}
\newcommand{\qq}{\vartheta}

\newcommand{\s}{\sigma}

\newcommand{\w}{\omega}

\newcommand{\z}{\zeta}

\newcommand{\W}{\Omega}

\renewcommand{\aa}{\mathcal{A}}
\newcommand{\bb}{\mathcal{B}}

\newcommand{\vc}[1]{\mathbf{#1}}
\newcommand{\tl}[1]{\tilde{#1}}

\newcommand{\avg}[1]{\left\langle #1 \right\rangle}

\newcommand{\mat}[2]{\left[ \begin{array}{#1} #2 \end{array} \right]}

\newcommand{\comments}[1]{}

\begin{document}

\title{Supersolid phases of dipolar bosons in optical lattices with a staggered flux}

\author{O. Tieleman}
\email{o.tieleman@uu.nl}

\affiliation{Institute for Theoretical Physics, Utrecht
University, Leuvenlaan 4, 3584 CE Utrecht, The Netherlands}

\author{A. Lazarides}

\affiliation{Institute for Theoretical Physics, Utrecht
University, Leuvenlaan 4, 3584 CE Utrecht, The Netherlands}

\author{C. Morais Smith}

\affiliation{Institute for Theoretical Physics, Utrecht
University, Leuvenlaan 4, 3584 CE Utrecht, The Netherlands}

\date{\today}


\begin{abstract}
We present the theoretical mean-field zero-temperature phase diagram of a Bose-Einstein condensate (BEC) with dipolar interactions loaded into an optical lattice with a staggered flux. Apart from uniform superfluid, checkerboard supersolid and striped supersolid phases, we identify several supersolid phases with staggered vortices, which can be seen as combinations of supersolid phases found in earlier work on dipolar BECs and a staggered-vortex phase found for bosons in optical lattices with staggered flux. By allowing for different phases and densities on each of the four sites of the elementary plaquette, more complex phase patterns are found.
\end{abstract}
\vskip2pc

\maketitle

\section{Introduction}

During the last decade, the field of ultracold atomic gases in optical lattices has undergone tremendous growth \cite{bloch08}, having witnessed amongst others the realisation of the superfluid-Mott insulator (SF-MI) transition \cite{misf2d, misf3d}, the construction of spin-dependent lattices \cite{spindeplatt}, and the generation of synthetic gauge fields \cite{synthgf}. In the earlier years, substantial effort was aimed at realising and understanding the Bose- and Fermi-Hubbard models, which feature only on-site interactions. Recently, there has been much progress in bringing longer-ranged interactions into the system by using dipolar atoms or molecules \cite{lahaye09}. An example of experimental progress in Bose-Einstein condensation of dipolar atoms can be found in Ref.~\cite{gries05}. Another approach is the creation of dipolar molecules; see e.g. Ref.~\cite{ni10}. Theoretical studies have indicated numerous interesting possibilities, including, but not limited to, supersolidity \cite{lahaye09, sengupta05, sdm09, tref09}.


Supersolidity is commonly defined as the simultaneous presence of diagonal and off-diagonal long-range order in the system \cite{chester70}, a prominent candidate for experimental realisation being solid $^4$He \cite{kim04, prokof07}. Other candidates have been suggested in the domain of ultracold atomic gases, such as rapidly rotating Fermi-Fermi mixtures \cite{moller07} and dipolar bosonic gases in optical lattices \cite{sdm09}. In an optical lattice, one clearly has diagonal long-range order, since the density at the minima of the lattice potential is higher than at the maxima; however, this type of long-range order is imposed externally. To preserve the analogy to bulk supersolids, where the diagonal order is spontaneously present in the system, we define a supersolid in an optical lattice as a phase with both long-range off-diagonal and diagonal order, where the diagonal order breaks the translational symmetry of the lattice \cite{sen08}. Recently, dipolar atoms or molecules in optical lattices have been predicted to feature such supersolidity. Refs.~\cite{tref09, pollet10} present analytical and numerical analyses of dipolar atoms in square lattices with only nearest-neighbour (NN) hopping. Other examples include square lattices with NN and next-nearest-neighbour hopping \cite{schmidt08} and triangular lattices \cite{sen08}.

Uniform magnetic fields for ultracold atomic gases have been mimicked by applying rotation \cite{cooper08, moller07} and by generating gauge fields using position-dependent optical coupling between internal states of the atoms \cite{lin09, juzeliunas06}. Analogous to superconductors in magnetic fields, these systems exhibit vortices. A staggered gauge field for neutral atoms has also been proposed \cite{ah07}, leading to a staggered-vortex superfluid phase \cite{lk10}. In this paper, we find that a dipolar bosonic gas subjected to a staggered gauge field exhibits a supersolid phase which features vortices. In contrast to Ref.~\cite{moller07}, we study the gas in a lattice and do not have a rotating trap.

We analyse the interplay between NN interactions and an artificial staggered magnetic field in a system of bosons in a two-dimensional square optical lattice. In order to perform this analysis, we generalise and combine the methods used in Refs.~\cite{sdm09}, \cite{lk10}, and \cite{lk08}, to allow for the description of phases with a higher degree of broken symmetry than discussed in those three references. We present a phase diagram containing combinations of the uniform and staggered-vortex phases found in Ref.~\cite{lk10} and the supersolid phases found in Ref.~\cite{sdm09}, as well as a region where two phases coexist and the system will phase separate. We find that several continuous and discontinuous phase transitions between different superfluid and supersolid phases can be driven in two ways: by changing the NN interaction strength or by changing the applied flux. Apart from the presence or absence of density modulations, we discuss the existence of another type of structure in the system, which arises when the many-body wavefunction exhibits phase differences between neighbouring lattice sites. The vortices studied in Refs.~\cite{lk10} and \cite{lk08} are a realisation of a non-trivial, although relatively simple phase structure.

This paper is structured as follows: In section \ref{secsystem}, we introduce the system and briefly discuss its constituent components. In section \ref{secmethod}, we present the methods used to determine the phase diagram, which is displayed and discussed in sections \ref{secsymm} and \ref{secasymm}. In section \ref{secexpsig} we show experimental signatures of the phases found. Section \ref{secconclude} concludes the paper by summarizing and discussing the results.

\section{The system}\label{secsystem}
We consider a system of dipolar bosons in a two-dimensional square optical lattice \cite{sdm09} with staggered flux \cite{lk08}. Below, we briefly explain the consequences of a staggered flux (section \ref{secflux}) and dipolar interactions in a lattice (section \ref{secdipole}) for the phases found in the system. We work at $T = 0$ and only consider Bose-condensed phases, since we are interested in the combined effects of a staggered flux and anisotropic NN interaction in a superfluid. The Hamiltonian we investigate is
\begin{align}
\begin{split}
\label{eqrealspaceham}
H = & \, H_{\rm flux} + H_{\rm on-site} + H_{\rm dip} \\
= & \, - J \sum_{\vc{r} \in \aa, l} \Bigl( e^{i (-1)^l \f} a^\dag_\vc{r} b_{\vc{r} + \vc{e}_l} + \mbox{H.c.} \Bigr) \\
& \, + \frac{U}{2} \sum_{\vc{r} \in \aa \oplus \bb} n_\vc{r} (n_\vc{r} - 1) \\
& \, + \sum_{\vc{r} \in \aa} \sum_{\s = \pm 1} \Bigl( V_x n_\vc{r} n_{\vc{r} + \s \vc{e}_1} + V_y n_\vc{r} n_{\vc{r} + \s \vc{e}_2} \Bigr).
\end{split}
\end{align}
Here, $J$ represents the hopping strength between neighboring sites; $\f$ is the phase picked up at each jump, which is related to the magnitude of the flux through a plaquette; $U$ the on-site interaction strength; and $V_x$ and $V_y$ the anisotropic NN interaction strengths. The lattice is represented as two interspersed square sublattices, $\aa$ and $\bb$. The operators $a_\vc{r}$ ($a^\dag_\vc{r}$) and $b_\vc{r'}$ ($b^\dag_\vc{r'}$) are destruction (creation) operators acting on sites in the sublattices $\aa$ and $\bb$, respectively; note that there is only one type of particle being created and destroyed. The operator $n_\vc{r}$ is the number operator for site $\vc{r}$, irrespective of the sublattice in which it is located. The lattice vectors $\vc{e}_l$, $l \in \{1, 2, 3, 4\}$ are defined by $\vc{e}_1 = - \vc{e}_3 = \vc{e}_x$ and $\vc{e}_2 = - \vc{e}_4 = \vc{e}_y$.

\subsection{Staggered flux}\label{secflux}
The term $H_{\rm flux}$ breaks the symmetry between the sublattices $\aa$ and $\bb$, as can be seen in Fig.~1(a). It can be represented as a synthetic magnetic field which alternates in sign between neighboring plaquettes. For the details of the derivation of this hopping term, we refer the reader to Refs.~\cite{lk10} and \cite{ah07}. The phase diagram of bosons with on-site interactions in such a lattice is presented in Ref.~\cite{lk10}; we reproduce it in Fig.~1(b).
\begin{figure}[h]\label{figfluxpd}
\begin{center}
\includegraphics[width=.25\textwidth]{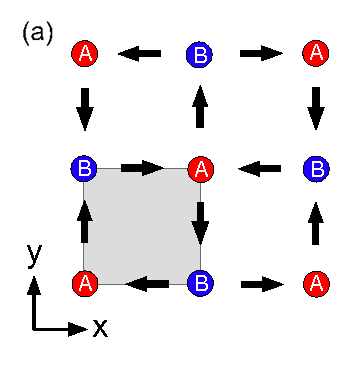}
\hspace{5pt}
\includegraphics[width=.16\textwidth]{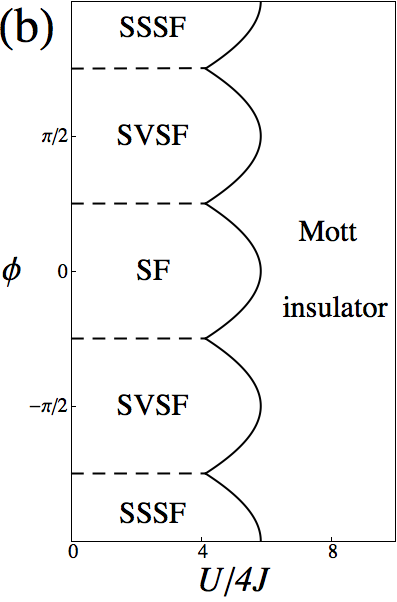}
\hspace{5pt}
\caption{(color online). (a) The division into sublattices $\aa$ and $\bb$ caused by the staggered flux. (b) The phase diagram for bosons with on-site interactions in an optical lattice with staggered flux as found by the authors of Ref.~\cite{lk10}. Legend: SF = conventional superfluid; SVSF = staggered-vortex superfluid; SSSF = staggered-sign superfluid.}
\end{center}
\end{figure}
The main conclusions from Ref.~\cite{lk10} which will be important for this paper are the following: For strong on-site interactions (large $U$), a Mott-insulating phase is found. By reducing $U$, a second-order phase transition into a superfluid phase occurs at some critical value of $U / J$; the value of $(U / J)_c$ depends on $\f$ as shown in Fig.~1(b). For $|\f|< \p / 4$, {\it i.e.}~small flux, the zero-momentum superfluid phase is unaltered; for $\p / 4 < |\f| < 3 \p / 4$, the system features a staggered-vortex superfluid phase, where the vortex cores are located at the centers of the plaquettes and the sign of the vorticity alternates between plaquettes. This phase comes about due to the development of a second minimum in the single-particle spectrum, at a finite momentum, which becomes the global minimum if $\p / 4 < |\f| < 3 \p / 4$. Condensation in this minimum introduces phase differences of $\p / 2$ between the lattice points, in such a pattern that a particle tunnelling around a plaquette picks up a phase of $\pm 2 \p$, depending on the direction of tunnelling and the plaquette. Note the periodicity in the $\f$-dependence of $(U/J)_c$. We find that the same periodic pattern emerges in the $V_x$-$\f$-diagrams that we present in sections \ref{secsymm} and \ref{secasymm}, respectively, in spite of the fact that we are studying different phase transitions. We will confine ourselves to the weakly interacting limit, since our aim here is to study the interplay between the supersolidity found in Ref.~\cite{sdm09} and the staggered-vortex patterns found in Ref.~\cite{lk10}.

\subsection{Dipolar interaction}\label{secdipole}
The interaction energy between two polarized dipoles is given by
\begin{align*}
v_{dd}(\vc{r}) = d^2 g_{dd} \frac{1 - 3 \cos^2 \z}{r^3},
\end{align*}
where $\z$ is the angle between the polarisation axis and the displacement vector $\vc{r}$. Loading the dipoles into a deep lattice and approximating the dipolar interaction by cutting it off at NN distance, the only displacement vectors that we have to consider are $\vc{e}_1$ and $\vc{e}_2$, the lattice vectors in the $x$- and $y$-directions. The NN dipolar interaction strengths in the two relevant directions are given by
\begin{align*}
V_x = & \, d^2 g_{dd} \frac{1 - 3 \sin^2 \qq \cos^2 \ff}{a^3}, \\
V_y = & \, d^2 g_{dd} \frac{1 - 3 \sin^2 \qq \sin^2 \ff}{a^3},
\end{align*}
where $a$ is the lattice spacing, $\qq$ is the inclination, and $\ff$ is the azimuthal angle. At $\ff = \p / 4$, the interaction strength is isotropic and can be varied continuously from repulsive ($\qq = 0$) to attractive ($\qq = \p / 2$), being zero at $\qq = \sin^{-1} \sqrt{2/3}$. By varying the azimuthal angle, we can tune the ratio between $V_x$ and $V_y$ to any desired value. We note that tuning the NN interactions to zero will make the next-nearest-neighbor interactions more relevant; however, in this paper, we focus on the strong NN interaction regime. Since the relevant quantities for the purposes of our analysis are $V_x / U$ and $V_y / U$, we can cover the complete $V_x/U$-$V_y/U$-plane by tuning $U$, $\q$ and $\ff$. Thus, we simply have two independent dimensionless parameters, $V_x/U$ and $V_y/U$. The Hamiltonian for the dipolar interaction thus takes the form
\begin{align}
H_{\rm dip} = \frac{1}{2} \sum_{\vc{r}, \s = \pm 1} \Bigl( V_x n_\vc{r} n_{\vc{r} + \s \vc{e}_1} + V_y n_\vc{r} n_{\vc{r} + \s \vc{e}_2} \Bigr).
\end{align}

In Ref.~\cite{sdm09}, the phase diagram of dipolar bosons in a square optical lattice is presented, which we reproduce in Fig.~2 and briefly discuss here. At mean-field-level, the authors identify three types of superfluid phases: a conventional superfluid one (SF), for weak NN interactions; one with a density modulation in a checkerboard pattern (checkerboard supersolid - CSS), for strong enough repulsive $V_x \approx V_y$; and one with a striped pattern (striped supersolid - SSS), where only one of the NN interaction parameters dominates. Both the checkerboard and striped superfluids have long-range diagonal as well as off-diagonal order, {\it i.e.}~ they break both gauge and translational symmetry. These properties justify the epithet supersolid, as discussed in the introduction.
\begin{figure}[h]
\begin{center}
\includegraphics[width=.4\textwidth]{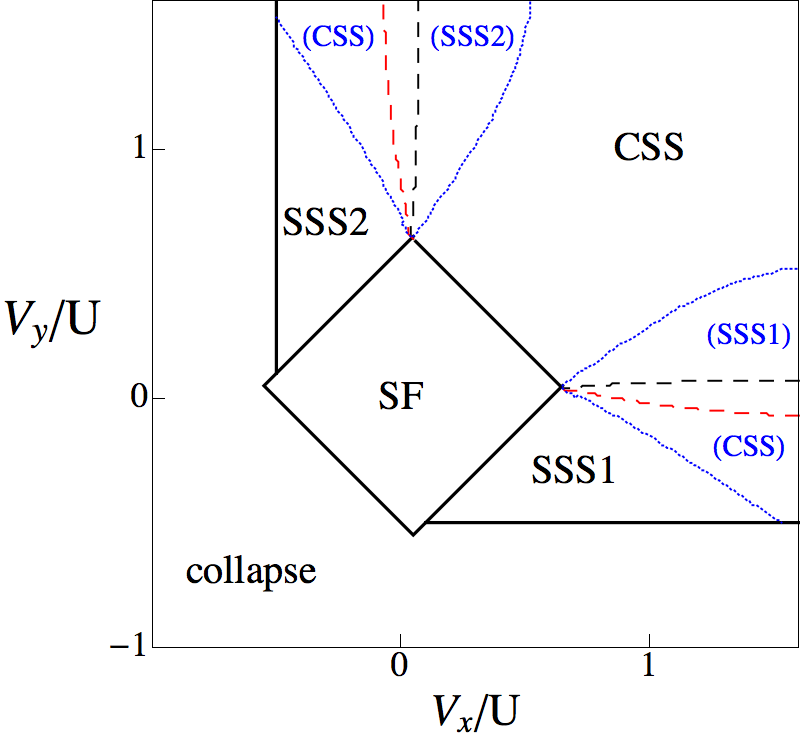}
\caption{(color online) The phase diagram for dipolar bosons in a square optical lattice. The black and blue lines were also found by the authors of Ref.~\cite{sdm09}. Legend: SF = homogeneous superfluid, CSS = checkerboard supersolid, SSS = striped supersolid. SSS1 (SSS2) has the stripes in the $y$-($x$-)direction. The solid black lines are second-order phase boundaries, the dashed black lines first-order phase transitions, and the dotted blue (grey) lines represent the existence of metastable states, which are labelled in blue (grey) and  between brackets. The red (grey) dashed lines are our finding, and represent the SSS-CSS phase boundary for the case of strong flux, $\f = \p / 2$; see the discussion in section \ref{secasymm}.}
\end{center}
\end{figure}

\noindent The checkerboard and striped phases are intuitively easy to understand. Consider the case where $V_x = V_y$: here, the NN interaction energy is reduced by arranging the atoms in a checkerboard-modulated density pattern, since there are fewer pairs of nearest neighbors in such a configuration. Similarly, if $V_y$ is negative (or positive but small) and $V_x$ is positive and of the order of $U$, the NN interaction energy is reduced by a striped configuration. As a final remark, taking longer-range interactions into account does affect the phase diagram to some extent, but the basic structure of checkerboard and striped phases remains intact~\cite{sdm09}.

\section{The method: mean-field}\label{secmethod}
Below, we explain how we obtain the phase diagram that we present in sections \ref{secsymm} and \ref{secasymm}. We use the Bogolyubov approximation to describe the condensate, which is justified if $J \gg U, V_x, V_y$ \cite{sdm09}; we calculate the ground state energies and excitation spectra of all the phases we identify. The excitation spectrum is required to check the dynamical stability of the phases we find: since we are dealing with bosons, a negative or complex excitation spectrum implies that the ground state above which the spectrum is calculated is not the real ground state of the system.

We will be working in the limit of high filling factor $\n$ and strong hopping (or weak interaction), such that we can consider every lattice point to contain a condensate, and still have a small $J / (\n U)$, where $\n$ is the filling factor. The ratio $J / (\n U)$ is required to be small for the density modulations to appear; if the energy gain from wavefunction overlaps between neighboring sites is too large, the system will ignore the NN interactions and simply remain in a superfluid state with homogeneous density.

\subsection{Two-sublattice formalism}
As described in section \ref{secflux}, the staggered flux divides the square lattice into two interpenetrating square sublattices, necessitating a two-sublattice description of the system, which is developed below. In the next subsection, we introduce another subdivision, into four sublattices, in order to allow for more complex density and phase distributions around the elementary plaquette.

We will perform our calculations in momentum space, where the grand-canonical version of the Hamiltonian given in Eq.~\eqref{eqrealspaceham} reads
\begin{align}
\begin{split}
\label{eqmomham}
H - \m N = & \, \sum_{\vc{k} \in {\rm BZ}_1} \Bigl( \e_\vc{k} a^\dag_\vc{k} b_\vc{k} + {\rm h.c.} - \m a^\dag_\vc{k} a_\vc{k} - \m b^\dag_\vc{k} b_\vc{k} \Bigr) \\
+ & \, \sum_{\substack{\vc{k}_1, ..., \vc{k}_4 \\ \in {\rm BZ}_1}} \delta_{\vc{k_1} + \vc{k_2} - \vc{k_3} - \vc{k_4}, 0} \\
\times & \, \biggl[ \frac{U}{2 N_{\rm s}} \Bigl( a^\dag_{\vc{k}_1} a^\dag_{\vc{k}_2} a_{\vc{k}_3} a_{\vc{k}_4} + b^\dag_{\vc{k}_1} b^\dag_{\vc{k}_2} b_{\vc{k}_3} b_{\vc{k}_4} \Bigr) \\
& \qquad \quad \; + \frac{2 V(\vc{k}_4 - \vc{k}_2)}{N_{\rm s}} a^\dag_\vc{k_1} b^\dag_\vc{k_2} a_\vc{k_3} b_\vc{k_4} \biggr],
\end{split}
\end{align}
with
\begin{align*}
\e_\vc{k} = - 4 J \biggl[ & \, \cos(\f) \cos \biggl( \frac{k_x + k_y}{2} a \biggr) \cos \biggl( \frac{k_x - k_y}{2} a \biggr) \\
+ & \, i \sin(\f) \sin \biggl( \frac{k_x + k_y}{2} a \biggr) \sin \biggl( \frac{k_x - k_y}{2} a \biggr) \biggr],
\end{align*}
and
\begin{align*}
V(\vc{k}) = V_x \cos(\vc{e}_x \cdot \vc{k}) + V_y \cos(\vc{e}_y \cdot \vc{k}).
\end{align*}
$N$ is the particle number operator for the entire system, $\m$ is the chemical potential, $N_{\rm s}$ is the number of sites per sublattice, {\it i.e.}~half the total number of sites, and BZ$_1$ is the first Brillouin zone. Now, we apply the Bogolyubov approximation: we replace $a_\vc{k} \to \delta_{\vc{k}, \vc{c}} \avg{a} + \tl{a}_\vc{k}$ and $b_\vc{k} \to \delta_{\vc{k}, \vc{c}} \avg{b} + \tl{b}_\vc{k}$, where $\vc{c}$ is the condensation momentum, and we treat $\tl{a}_\vc{k}$ and $\tl{b}_\vc{k}$ as small fluctuations relative to the average occupations $\avg{a}$ and $\avg{b}$. We then require that the terms linear in the fluctuations vanish. This requirement yields the values for the chemical potential and, if present, the density modulation; for the details, see sections \ref{secsymm} and \ref{secasymm}. The terms of order zero in the fluctuations represent the ground state energy,
\begin{align}
\begin{split}
E_0 = & \, 2 {\rm Re} (\e_\vc{c} \avg{a} \avg{b}^*) + U (|\avg{a}|^4 + |\avg{b}|^4) / 2 N_{\rm s} \\
& \, + 2 V(\vc{0}) |\avg{a}|^2 |\avg{b}|^2 / N_{\rm s}.
\end{split}
\end{align}
The second-order terms can be diagonalised to give the excitation spectrum. They can be represented in matrix-form as
\begin{align}\label{eqspec2sl}
H^{\rm ex} = & \, \frac{1}{2} \sum_{\vc{k} \in {\rm BZ}_1} A^\dag_\vc{k}
\mat{cccc}{
\w_\vc{k} & \lambda_\vc{k} & \g_\vc{k} & \z_\vc{k} \\
\lambda^*_\vc{k} & \w_\vc{k} & \z^*_\vc{k} & \g^*_\vc{k} \\
\g^*_\vc{k} & \z_\vc{k} & \w_\vc{k} & \lambda_\vc{k} \\
\z^*_\vc{k} & \g_\vc{k} & \lambda^*_\vc{k} & \w_\vc{k}} A_\vc{k} \notag \\
= & \, \frac{1}{2} \sum_{\vc{k} \in {\rm BZ}_1} A^\dag_\vc{k} M_\vc{k} A_\vc{k},
\end{align}
where
\begin{align*}
A^\dag_\vc{k} = \mat{cccc}{a^\dag_\vc{k} & a_\vc{-k} & b^\dag_\vc{k} & b_\vc{-k}},
\end{align*}
and $\w_\vc{k}$, $\lambda_\vc{k}$, $\g_\vc{k}$, and $\z_\vc{k}$ are functions to be calculated for each specific phase we describe with this formalism. We diagonalise this quadratic Hamiltonian by solving
\begin{align*}
\Bigl( M_\vc{k} - \W_\vc{k} [A_\vc{k}, A^\dag_\vc{k}] \Bigr) = 0,
\end{align*}
for $\W_\vc{k}$, which then is the excitation spectrum.

\subsection{Four-sublattice description}\label{sec4sl}
The method presented above is in fact a simple combination of the approaches used in Refs.~\cite{sdm09} and \cite{lk10}. It works as long as either the phase distribution around the elementary plaquette is trivial ({\it i.e.}~all sites have the same phase) or the density modulation is absent. If we allow $V_x \neq V_y$, striped or otherwise asymmetric phases may occur, in which case the four sites of the elementary plaquette could all have different densities, or the phase drops could be distributed unevenly along the plaquette. In such phases, the two-sublattice formalism does not hold, since the density modulation and the phase pattern will influence each other, as we show below. One interesting phenomenon to be investigated here is the competition between NN interactions favouring stripes on the one hand, and a staggered flux on the other hand, since the staggered flux is associated with checkerboard-subdivision of the lattice (see section \ref{secflux}, Fig.~1). To investigate such phenomena, we need a description of the system which allows the condensate wavefunction to be different at all four sites of the elementary plaquette. Such a description involves four different sublattices: we have to split each of the sublattices $\aa$ and $\bb$ into two new ones, such that $\aa = $ SL$_1 \oplus$SL$_3$ and $\bb = $SL$_2 \oplus$SL$_4$:
\begin{align*}
\begin{array}{cccc}
a & b & a & b \\
b & a & b & a
\end{array} \to 
\begin{array}{cccc}
1 & 2 & 1 & 2 \\
4 & 3 & 4 & 3
\end{array}.
\end{align*}
The resulting equations for the chemical potential and condensate wavefunction cannot be solved analytically, and have to be solved numerically instead. In this representation, the momentum-space Hamiltonian becomes
\begin{widetext}
\begin{align}
\begin{split}
\label{eqmomham4sl}
H = & \, \sum_{\vc{k} \in {\rm BZ}_1} \Bigl( \e^x_\vc{k} a^\dag_{1,\vc{k}} a_{2,\vc{k}} + (\e^y_\vc{k})^* a^\dag_{1,\vc{k}} a_{4,\vc{k}} + \e^x_\vc{k} a^\dag_{3,\vc{k}} a_{4,\vc{k}} + (\e^y_\vc{k})^* a^\dag_{3,\vc{k}} a_{2,\vc{k}} \Bigr) + {\rm h.c.} \\
- & \, \m \sum_{\vc{k} \in {\rm BZ}_1} \sum_{i = 1}^4 a^\dag_{i, \vc{k}} a_{i, \vc{k}} + \sum_{\vc{k}_1, ..., \vc{k}_4} \delta_{\vc{k}_1 + \vc{k}_2 - \vc{k}_3 - \vc{k}_4, 0} \biggl[ \frac{U}{2 N} \sum_{i = 1}^4 a^\dag_{i,\vc{k}_1} a^\dag_{i,\vc{k}_2} a_{i,\vc{k}_3} a_{i,\vc{k}_4} \\
+ & \, \frac{2 V_x}{N} \cos[(\vc{k}_4 - \vc{k}_2) \cdot \vc{e}_x] \Bigl( a^\dag_{1,\vc{k}_1} a^\dag_{2,\vc{k}_2} a_{1,\vc{k}_3} a_{2,\vc{k}_4} + a^\dag_{3,\vc{k}_1} a^\dag_{4,\vc{k}_2} a_{3,\vc{k}_3} a_{4,\vc{k}_4} \Bigr) \\
+ & \, \frac{2 V_y}{N} \cos[(\vc{k}_4 - \vc{k}_2) \cdot \vc{e}_y] \Bigl( a^\dag_{1,\vc{k}_1} a^\dag_{4,\vc{k}_2} a_{1,\vc{k}_3} a_{4,\vc{k}_4} + a^\dag_{2,\vc{k}_1} a^\dag_{3,\vc{k}_2} a_{2,\vc{k}_3} a_{3,\vc{k}_4} \Bigr) \biggr],
\end{split}
\end{align}
\end{widetext}
where
\begin{align*}
e^{x/y}_\vc{k} = & \, - 2 J \cos(k_{x/y} a) e^{i \f}.
\end{align*}
Note that this four-sublattice formalism can in principle be used to describe all the phases that we will encounter in the system, including the ones that can be analysed within the two-sublattice formalism; for example, the checkerboard-modulated phase without vortices has $\avg{a_1} = \avg{a_3} = \beta \avg{a_2} = \beta \avg{a_4}$, where $\beta$ is a measure for the strenght of the modulation.

\comments{$\avg{a_1} = \avg{a_3} = \beta \avg{a_2} = \beta \avg{a_4}$}
\comments{$\langle a_1 \rangle = \langle a_3 \rangle = \beta \langle a_2 \rangle = \beta \langle a_4 \rangle$}

\subsubsection{Mean field}
In the four-sublattice description, the real space unit cell is twice as large as in the two-sublattice description. As a consequence, the first Brillouin zone is only half the size, and the number of bands in the excitation spectrum is doubled. The minimum in the corner of the first Brillouin zone found in Ref.~\cite{lk10} is mapped to the center of the new first Brillouin zone. Hence, the minimum of the single-particle spectrum in the four-sublattice description is always at $\vc{k} = 0$, and we use the mean-field ansatz $a_{j, \vc{k}} \to \delta_{\vc{k}, 0} \avg{a_j} + \tl{a}_{j, \vc{k}}$. We obtain four equations for the chemical potential, of the form
\begin{align}\label{eqmu1}
\begin{split}
\m_1 = & \, - 2 J \frac{e^{i \f} \avg{a_2} + e^{- i \f} \avg{a_4}}{\avg{a_1}} + U \n |\avg{a_1}|^2 \\
+ & \, 2 V_x \n |\avg{a_2}|^2 + 2 V_y \n |\avg{a_4}|^2
\end{split}
\end{align}
(we omit the other three for brevity; they can easily be deduced from the Hamiltonian). These four $\m_j$ can be interpreted as the chemical potentials of the four sublattices, which are then thought of as macroscopic systems with an exchange mechanism (the hopping terms). The condition that all four chemical potentials are equal represents the equilibrium condition in this picture. Since the expression in Eq.~\eqref{eqmu1} is in principle complex, we have to allow for complex $\avg{a_j}$ in order to be able to make the imaginary parts of the $\m_j$ vanish. Representing $\avg{a_j}$ as $r_j e^{i \q_j}$, the requirement that all $\m_j$ are real yields the conditions
\begin{align}
\begin{split}
\label{eqchempotim}
r_2 \sin \a_1 = & \, r_4 \sin \a_4 \\
r_3 \sin \a_2 = & \, r_1 \sin \a_1 \\
r_4 \sin \a_3 = & \, r_2 \sin \a_2 \\
r_1 \sin \a_4 = & \, r_3 \sin \a_3,
\end{split}
\end{align}
where $\a_j = \q_{j+1} - \q_j + \f$. The four equations \eqref{eqchempotim} can be reduced to three without loss of generality. A fourth equation comes from the requirement that the mean-field wavefunction is always single-valued. To satisfy this requirement, the phase picked up when hopping around a plaquette has to be an integer multiple of $2 \p$; thus,
\begin{align}\label{eqsingvalwf}
\sum_i \a_i - 4 \f = 2 \p n,
\end{align}
where $n$ determines the vorticity pattern of the system. Apart from Eqs.~\eqref{eqchempotim}, we also have the real parts of the chemical potentials, which have to be equal to each other. They take the form
\begin{align}
\begin{split}\label{eqchempotreal}
\m_1 = & \, - \frac{2 J}{r_1} \bigl( r_2 \cos \a_1 + r_4 \cos \a_4 \bigr) + U \n r_1^2 \\
& \, + 2 V_x \n r_2^2 + 2 V_y \n r_4^2,
\end{split}
\end{align}
and similar expressions for $\m_2$, $\m_3$ and $\m_4$. Finally, there is the normalisation condition,
\begin{align}\label{eqdensnorm}
r_1^2 + r_2^2 + r_3^2 + r_4^2 = 4,
\end{align}
bringing us to a total of eight equations in eight unknowns. Of course, more than one unique solution may still exist. We have to find all unique solutions, since each represents a different phase of the system, and we want to compare all phases for stability and ground state energy. Hence, we solve the eight equations Eqs.~\eqref{eqchempotim}-\eqref{eqdensnorm} numerically, in such a way that we find all solutions. Having found the $\avg{a_j}$, the ground state energy is given by
\begin{align}\label{eqgse4sl}
\begin{split}
E_0 = & \, - 4 J \, {\rm Re} \, \sum_j e^{i \f} \avg{a_j}^* \avg{a_{j+1}} + \frac{U}{2 N} \sum_j |\avg{a_j}|^4 \\
& \, + \frac{2 V_x}{N_{\rm s}} \bigl( |\avg{a_1}|^2 |\avg{a_2}|^2 + |\avg{a_3}|^2 |\avg{a_4}|^2 \bigr) \\
& \, + \frac{2 V_y}{N_{\rm s}} \bigl( |\avg{a_1}|^2 |\avg{a_4}|^2 + |\avg{a_3}|^2 |\avg{a_2}|^2 \bigr),
\end{split}
\end{align}
where $j$ is to be taken modulo $4$.

\subsubsection{Excitation spectrum}
As before, finding all solutions that represent equilibrium situations is not enough: we also have to investigate their excitation spectra to assess their respective dynamical stabilities. In order to derive the excitation spectra, we collect all terms of second order in the fluctuations, and obtain, quite generally,
\begin{align}\label{eqspec4sl}
\begin{split}
H^{\rm 4sl}_{\rm ex} = & \, \frac{1}{2} \sum_{\vc{k} \in {\rm BZ}_1} \sum_{i, j} \Bigl[ \w^{ij}_\vc{k} a^\dag_{i, \vc{k}} a_{j, \vc{k}} + \w^{ji}_\vc{k} a^\dag_{j, \vc{k}} a_{i, \vc{k}} \\
+ & \, (\lambda^{ij}_\vc{k})^* a^\dag_{i, \vc{k}} a^\dag_{j, -\vc{k}} + \lambda^{ij}_\vc{k} a_{i, \vc{k}} a_{j, -\vc{k}} \Bigr],
\end{split}
\end{align}
where
\begin{align*}
A^\dag_\vc{k} = \mat{cccccccc}{a^\dag_{1, \vc{k}} & a_{1, -\vc{k}} & a^\dag_{2, \vc{k}} & a_{2, -\vc{k}} & a^\dag_{3, \vc{k}} & a_{3, -\vc{k}} & a^\dag_{4, \vc{k}} & a_{4, -\vc{k}}}.
\end{align*}
In our specific case, $\w^{ii}$ and $\lambda^{ii}$ do not depend on $\vc{k}$: they are given by
\begin{subequations}\label{eqmatelm4sl}
\begin{align}
\begin{split}
\w^{11}_\vc{k} = & \, 2 U |\avg{a_1}|^2 + 2 V_x |\avg{a_2}|^2 + 2 V_y |\avg{a_4}|^2 - \m \\
\w^{22}_\vc{k} = & \, 2 U |\avg{a_2}|^2 + 2 V_x |\avg{a_1}|^2 + 2 V_y |\avg{a_3}|^2 - \m \\
\w^{33}_\vc{k} = & \, 2 U |\avg{a_3}|^2 + 2 V_x |\avg{a_4}|^2 + 2 V_y |\avg{a_2}|^2 - \m \\
\w^{44}_\vc{k} = & \, 2 U |\avg{a_4}|^2 + 2 V_x |\avg{a_3}|^2 + 2 V_y |\avg{a_1}|^2 - \m \\
\lambda^{ii}_\vc{k} = & \, U (\avg{a_i}^*)^2.
\end{split}
\end{align}
The hopping terms have the following coefficients:
\begin{align}
\begin{split}
\w^{12}_\vc{k} = & \, (\w^{21}_\vc{k})^* = \e^x_\vc{k} + 2 V_x \cos(k_x a) \avg{a_1} \avg{a_2}^* \\
\w^{23}_\vc{k} = & \, (\w^{32}_\vc{k})^* = \e^y_\vc{k} + 2 V_y \cos(k_y a) \avg{a_2} \avg{a_3}^* \\
\w^{34}_\vc{k} = & \, (\w^{43}_\vc{k})^* = \e^x_\vc{k} + 2 V_x \cos(k_x a) \avg{a_3} \avg{a_4}^* \\
\w^{41}_\vc{k} = & \, (\w^{14}_\vc{k})^* = \e^y_\vc{k} + 2 V_y \cos(k_y a) \avg{a_4} \avg{a_1}^* \\
\w^{13}_\vc{k} = & \, (\w^{13}_\vc{k})^* = 0 \\
\w^{24}_\vc{k} = & \, (\w^{24}_\vc{k})^* = 0.
\end{split}
\end{align}
Lastly, the off-diagonal mixing terms read
\begin{align}
\begin{split}
\lambda^{12}_\vc{k} = & \, \lambda^{21}_\vc{k} = 2 V_x \cos(k_x a) \avg{a_1}^* \avg{a_2}^* \\
\lambda^{23}_\vc{k} = & \, \lambda^{32}_\vc{k} = 2 V_y \cos(k_y a) \avg{a_2}^* \avg{a_3}^* \\
\lambda^{34}_\vc{k} = & \, \lambda^{43}_\vc{k} = 2 V_x \cos(k_x a) \avg{a_3}^* \avg{a_4}^* \\
\lambda^{41}_\vc{k} = & \, \lambda^{14}_\vc{k} = 2 V_y \cos(k_y a) \avg{a_4}^* \avg{a_1}^* \\
\lambda^{13}_\vc{k} = & \, \lambda^{21}_\vc{k} = 0 \\
\lambda^{24}_\vc{k} = & \, \lambda^{21}_\vc{k} = 0.
\end{split}
\end{align}
\end{subequations}
As before, we diagonalise this Hamiltonian by numerically solving
\begin{align*}
\Bigl( M^{\rm 4sl}_\vc{k} - \W^{\rm 4sl}_\vc{k} [A_\vc{k}, A^\dag_\vc{k}] \Bigr) = 0
\end{align*}
for $\W^{\rm 4sl}_\vc{k}$, which then gives the excitation spectrum.

In most regions of parameter space, there is only one dynamically stable phase. However, there are some regions where two phases are dynamically stable; here, we compare the ground state energies (which are equal to the free energies, since we are working at $T = 0$) to see which phase is favoured.

The phase diagram we find is, in principle, three-dimensional, since we have the three parameters $\f$, $V_x$ and $V_y$. In addition to the $V_x / U$-$V_y / U$-diagram at $\f = 0$ from Ref.~\cite{sdm09}, we present three cross sections which together form a representative sample of the results: the $V_x$-$\f$-diagram at $V_y = V_x$, the $V_x$-$\f$-diagram at $V_y = 0$, and the $V_x/U$-$V_y/U$-diagram at $\f = \p / 2$.

\section{Quantum phases: symmetric case}\label{secsymm}
In this section, we discuss the symmetric case, where the NN interaction is equally strong in both directions. This can be achieved by polarising the dipoles perpendicular to the plane, or at an angle of $\p / 4$ relative to the in-plane lattice vectors. By tuning the inclination, the ratio $V_{x/y} / U$ can be tuned without changing $U$. However, if this technique is employed, the next-nearest-neighbor interactions will not be isotropic; this is only the case if the polarisation axis is perpendicular to the plane. As a consequence, the description presented here will be more accurate if the dipoles are polarised perpendicularly to the plane.

\subsection{Weak flux: no vortices}
For small values of the flux, $0 \leq |\f| \leq \p / 4$, the system does not feature any vortices in the ground state. It may still exhibit density modulations, since these are caused by the NN interactions.

\subsubsection{Superfluid: homogeneous density}
For homogeneous or checkerboard-modulated density distributions, we can use the two-sublattice formalism, and replace $a_\vc{k} \to \delta_{\vc{k}, 0} \avg{a} + \tl{a}_\vc{k}$ and $b_\vc{k} \to \delta_{\vc{k}, 0} \avg{b} + \tl{b}_\vc{k}$. If the density is homogeneous, we have a conventional superfluid (SF) and $|\avg{a}| = |\avg{b}| = \sqrt{N_{\rm p} / 2}$, with $N_{\rm p}$ being the total number of particles in the system. Setting the term linear in the fluctuations zero yields
\begin{align*}
\m_{\rm SF} = \e_0 + \n (U + 2 V_x + 2 V_y).
\end{align*}
Note that $\e_0 = - 4 J \cos \f$, and hence as $\f \to 0$, we recover the result found in Ref.~\cite{sdm09}: $\m_{\rm SF} = - 4 J + \n (U + 2 V_x + 2 V_y)$. In order to determine where this superfluid phase is dynamically stable, we consider the excitation spectrum given by Eq.~\eqref{eqspec2sl}. The matrix elements of $M^{\rm SF}_\vc{k}$ are
\begin{align*}
\w_\vc{k} = & \, 2 \n (U + V_x + V_y) - \m_{\rm SF} \\
\lambda_\vc{k} = & \, \n U \\
\g_\vc{k} = & \, \e_\vc{k} + \z_\vc{k} \\
\z_\vc{k} = & \, 2 \n V(\vc{k}).
\end{align*}

As $\f \to 0$, the excitation spectrum should reduce to $\W^{\rm SF}_\vc{k} = \sqrt{\tl{\e}_\vc{k} [\tl{\e}_\vc{k} + 2 \n (U + 2 V(\vc{k}))]}$, with $\tl{\e}_\vc{k} = 2 J [2 + \cos (k_x a) + \cos (k_y a)]$, as found in Ref.~\cite{sdm09}. However, in order to make this comparison properly, we have to map the two-band excitation spectrum in the Brillouin zone defined by $k_x \pm k_y \in [-\p / a, \p / a]$ derived above, to the single-band spectrum in the Brillouin zone $k_{x/y} \in [-\p / a, \p / a]$ given in Ref.~\cite{sdm09}. This mapping is described in e.~g.~Ref.~\cite{kittel}: it is the mapping from the extended to the reduced zone scheme. After applying this mapping, the two spectra are seen to be identical. In addition, as $V_x \to 0$ and $V_y \to 0$, the spectrum reduces to the one found in Ref.~\cite{lk10}.

\subsubsection{Checkerboard supersolid}
To describe a phase with a checkerboard-modulated density (checkerboard supersolid, CSS), we follow the scheme used in Ref.~\cite{sdm09}. We assume that the population density of sublattice $\aa$ is different from that of sublattice $\bb$; hence, $\avg{a} \neq \avg{b}$. Instead of working with $\avg{a}$ and $\avg{b}$ directly, we go to a center-of-mass and relative representation, and write
\begin{align*}
\avg{a} = & \, \sqrt{N_{\rm p} / 2} \bigl( \sqrt{\a} + \sqrt{\beta} \bigr) \\
\avg{b} = & \, \sqrt{N_{\rm p} / 2} \bigl( \sqrt{\a} - \sqrt{\beta} \bigr).
\end{align*}
Clearly, if we send $\beta$ to zero, the density modulation vanishes. Replacing $a_\vc{k} \to \delta_{\vc{k}, 0} \avg{a} + \tl{a}_\vc{k}$ and $b_\vc{k} \to \delta_{\vc{k}, 0} \avg{b} + \tl{b}_\vc{k}$ in Eq.~\eqref{eqmomham} and requiring the terms of first order in the fluctuations to vanish yields
\begin{align*}
\frac{\m_\aa}{\n} = & \, \frac{\e_\vc{0}}{\n} \frac{\sqrt{\a} - \sqrt{\beta}}{\sqrt{\a} + \sqrt{\beta}} + U + 2 V(0) + 2 \sqrt{\a \beta} [U - 2 V(0)], \\
\frac{\m_\bb}{\n} = & \, \frac{\e_\vc{0}}{\n} \frac{\sqrt{\a} + \sqrt{\beta}}{\sqrt{\a} - \sqrt{\beta}} + U + 2 V(0) - 2 \sqrt{\a \beta} [U - 2 V(0)],
\end{align*}
where $\n = N_{\rm p} / N_{\rm s}$ is the average number of atoms per site (filling factor) and we have used the fact that that $|\avg{a}|^2 + |\avg{b}|^2 = N_{\rm p}$ and hence $\a + \beta = 1$. Setting $\m_\aa = \m_\bb$ yields a condition on $\a$ and $\beta$, which can be solved for the difference $\a - \beta$; the result is
\begin{align}\label{eqcheckmod}
\a - \beta = \frac{4 J \cos \f}{\n (2 V_x + 2 V_y - U)}.
\end{align}
Now, we can now write $\a$ and $\beta$ in terms of the parameters $J$, $\f$, $U$, $V_x$ and $V_y$:
\begin{align*}
\a = & \, \frac{1}{2} + \frac{2 J \cos \f}{\n (2 V_x + 2 V_y - U)}, \\
\beta = & \, \frac{1}{2} - \frac{2 J \cos \f}{\n (2 V_x + 2 V_y - U)}.
\end{align*}
With this result, we can calculate the chemical potential, and find $\m_{\rm CSS} = 2 \n U$. For the checkerboard-modulated case, the matrix elements of $M^{\rm CSS}_\vc{k}$ in Eq.~\eqref{eqspec2sl} are given by
\begin{align*}
\w^{a/b}_\vc{k} = & \, 2 U \n \bigl( 1 \pm 2 \sqrt{\a \beta} \bigr) + 2 V(0) \n \bigl( 1 \mp 2 \sqrt{\a \beta} \bigr) - \m_{\rm CSS}, \\
\lambda^{a/b}_\vc{k} = & \, U \n \bigl( 1 + 2 \sqrt{\a \beta} \bigr), \\
\g_\vc{k} = & \, \e_\vc{k} + \z_\vc{k}, \\
\z_\vc{k} = & \, 2 \n V(\vc{k}).
\end{align*}
As in the homogeneous case, if we take the limit $\f \to 0$, we recover the results from Ref.~\cite{sdm09}: density modulation, chemical potential, and excitation spectrum. Note that for $0 < |\f| < \p / 4$, the density modulation is affected by the flux, even though no vortices appear (see Eq.~\eqref{eqcheckmod}).

\subsection{Strong flux: staggered-vortex phase}\label{secstrongflux}
As was found in Ref.~\cite{lk10}, under the influence of a strong flux, $\p / 4 < |\f| < 3 \p / 4$, the system goes to a staggered-vortex superfluid (SVSF) phase. This corresponds to condensation in the single-particle state with momentum $(\pm \p / a, 0)$ or $(0, \pm \p / a)$, {\it i.e.}~ in the corners of the first Brillouin zone \footnote{Note that although the condensation momenta do not look identical, they actually are, since the relevant reciprocal lattice vectors are those of the $\aa$ and $\bb$ sublattices, which do translate these condensation points into each other.}. To describe this region of parameter space, we need the four-sublattice formalism presented in section \ref{sec4sl}. Although the description is different, the {\it ansatz} is still informed by earlier findings: we expect a combination of the staggered-vortex pattern from Ref.~\cite{lk10} and, for appropriately strong NN interactions, the checkerboard-modulated density from Ref.~\cite{sdm09}. For a homogeneous density distribution, the {\it ansatz} is quite simple:
\begin{align*}
\avg{a_j} = \n e^{i (j - 1) \p / 2},
\end{align*}
where we have defined the phase of the mean-field wavefunction at sublattice SL$_1$ to be zero. For the discussion of the staggered-vortex checkerboard supersolid (SVCSS), there is a general point that is worth noting: in cases where the density modulation is invariant under exchange of the two lattice vectors, the vortices do not interfere with the density modulation. This can be seen from Eq.~\eqref{eqchempotim}: as long as $r_1 = r_3$ and $r_2 = r_4$, the conditions on the phase drops do not depend on the wavefunction amplitudes. Since the checkerboard pattern has this symmetry, we can try to guess the modulation strength from the two-sublattice formalism. The {\it ansatz} would be:
\begin{align*}
a_\vc{k} \to & \, \sqrt{N_{\rm p} / 2} \delta_{\vc{k}, \p} \bigl( \sqrt{\a} + \sqrt{\beta} \bigr) + \tl{a}_\vc{k} \\
b_\vc{k} \to & \, i \sqrt{N_{\rm p} / 2} \delta_{\vc{k}, \p} \bigl( \sqrt{\a} - \sqrt{\beta} \bigr) + \tl{b}_\vc{k}.
\end{align*}
Performing the same analysis as for the weak-flux case, we find that the density modulation strength is given by
\begin{align}\label{eqcheckmodsv}
\a - \beta = \frac{4 J \sin \f}{\n (2 V_x + 2 V_y - U)}
\end{align}
Hitherto, the two formalisms work equally well. However, the excitation spectrum can only be calculated in the four-sublattice formalism; the mean-field values for the four sublattices are
\begin{align*}
\avg{a_1} = & \, \sqrt{N_{\rm p} / 2} (\sqrt{\a} + \sqrt{\beta}) \\
\avg{a_2} = & \, - i \sqrt{N_{\rm p} / 2} (\sqrt{\a} - \sqrt{\beta}) \\
\avg{a_3} = & \, - \sqrt{N_{\rm p} / 2} (\sqrt{\a} + \sqrt{\beta}) \\
\avg{a_4} = & \, i \sqrt{N_{\rm p} / 2} (\sqrt{\a} - \sqrt{\beta}).
\end{align*}
By inserting these values for $\avg{a_j}$ into Eqs.~\eqref{eqgse4sl} and \eqref{eqmatelm4sl}, we find the corresponding ground state energy and excitation spectrum.

\subsection{Phase diagram}
Now we have all the information required to determine the cross section of the phase diagram along the line $V_x = V_y$. In order to obtain the phase boundaries, we calculate the parameter values where the density modulation vanishes, where the excitation spectra become unstable, and where the ground state energies of two phases are equal. These three phenomena happen simultaneously at the SF-CSS and SVSF-SVCSS phase boundaries (see Fig.~3). Since the relevant order parameter for this phase transition is the density modulation, which vanishes continously at the border, this is a second-order phase transition. At the SF-SVSF and CSS-SVCSS boundaries, the ground state energies of the uniform and staggered-vortex phases become equal. The relevant order parameter for these transitions is the (staggered) number of vortices per plaquette, which jumps from zero to unity; hence, these are first-order phase transitions.

Fig.~3 shows the $V_x = V_y$ cross section of the phase diagram. The combination of flux and NN interactions leads to a phase boundary with the same shape as observed in Ref.~\cite{lk10} in the MI-SF phase diagram. The phase transition from the homogeneous-density phase to the checkerboard phase is continuous, independently of the flux; the phase transition from the uniform to the staggered-vortex phase is discontinuous, independently of the density modulation.

In conclusion, this cross-section of the phase diagram shows a relatively straightforward superposition of the phases found in Refs.~\cite{sdm09} and \cite{lk10}. Beyond the results found in those references, we note a few aspects: The symmetry in the line $\f = \p / 4$ goes further than the shape of the phase boundary, since the strength of the density modulation is also symmetric. The origin of this symmetry can be found in Eqs.~\eqref{eqcheckmod} and \eqref{eqcheckmodsv}: the cosine from the uniform SF and CSS phases is replaced by a sine in the staggered-vortex phases. The SF-CSS transition can also be induced by changing $\f$, provided $V_x$ and $V_y$ are in the right range. The critical value of $V_x / U$ shifts with $J / \n U$, but the phase boundaries retain their periodicity in $\f$, within the approximation used here.

\begin{figure}[h]
\begin{center}
\includegraphics[width=.35\textwidth]{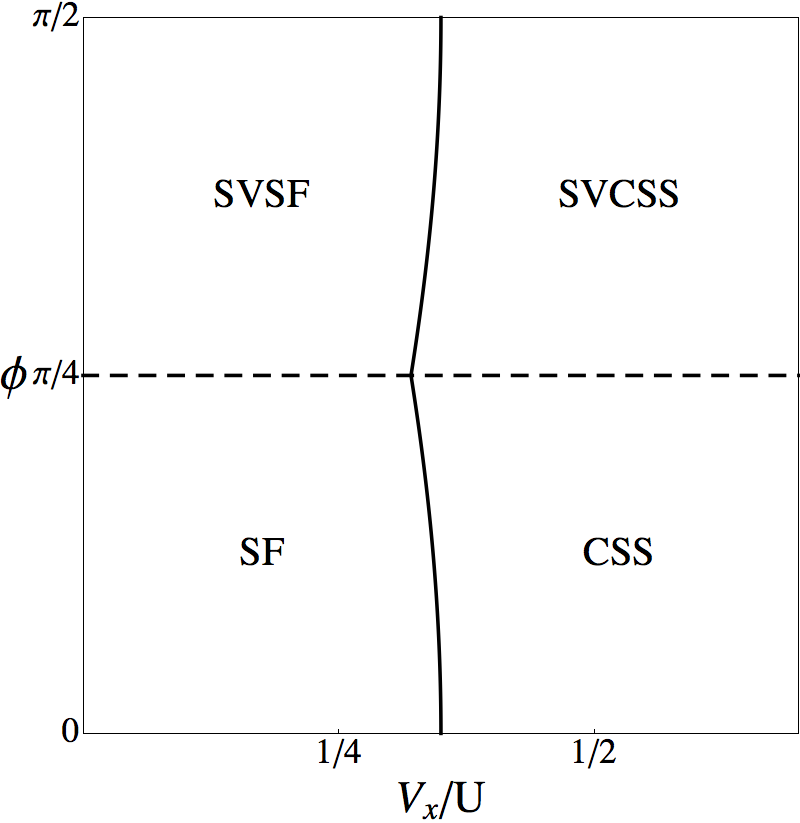}
\caption{Cross section of the phase diagram for $V_x = V_y$ and $J / (\n U) = 0.1$. Dashed (solid) lines are first (second) order phase transitions. Legend: SVSF = homogeneous staggered-vortex superfluid; SVCSS = staggered-vortex checkerboard supersolid.}
\end{center}
\end{figure}

\section{Quantum phases: asymmetric case}\label{secasymm}

In Eqs.~\eqref{eqchempotim} and \eqref{eqchempotreal}, we see the interplay between the phase and density distributions. As noted in section \ref{secstrongflux}, both a homogeneous density and a checkerboard pattern are possible within the {\it ansatz} employed in Ref.~\cite{lk10}, in which all phase drops are assumed to be equal. If, however, $r_1 \neq r_3$ and/or $r_2 \neq r_4$, as in the case of stripes or four different densities on the four sites of one plaquette, the phase distribution is influenced by the density modulation, and we have to allow for phase drops taking other values than integer multiples of $\p / 2$. We indeed find such solutions, in the striped phase for $0 < \f < \p / 2$. The mean-field wavefunction values $\avg{a_j}$ are determined by numerically solving Eqs.~\eqref{eqchempotim}-\eqref{eqdensnorm}.

\subsection{Phase diagram cross section I: $V_y = 0$}\label{secvy0}
We obtain the phase diagram in the same manner as in section \ref{secsymm}: by calculating where the density modulations vanish, where the excitation spectra develop instabilities, and by comparing ground state energies.
\begin{figure}[h]
\begin{center}
\includegraphics[width=.35\textwidth]{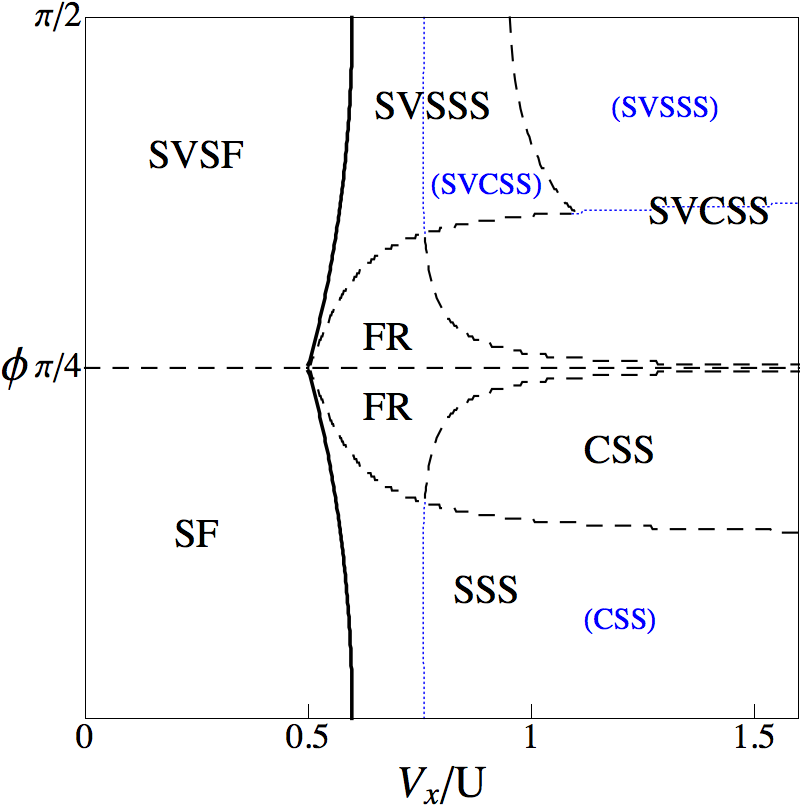}
\caption{Cross section of the phase diagram for $V_y = 0$ and $J / (\n U) = 0.1$. Legend: SSS = striped supersolid; SVSSS = staggered-vortex striped supersolid; FR = forbidden region, phase separation. Dotted blue (grey) lines represent the existence of metastable states, which are labelled in blue (grey) and between brackets.}
\end{center}
\end{figure}
Along the line $V_y = 0$, we find a variety of phases (see Fig.~4). The superposition of the CSS phase from Ref.~\cite{sdm09} with the staggered-vortex phase from Ref.~\cite{lk10} is there, as well as the uniform and staggered-vortex superfluids without density modulation. The striped supersolid (SSS) and staggered-vortex striped supersolid (SVSSS) phases are not simple combinations of earlier found phases, however: they break the symmetry within the sublattices $\aa$ and $\bb$, {\it i.e.}~ between sublattices SL$_1$ and SL$_3$, and SL$_2$ and SL$_4$, respectively. The density modulates in a striped pattern, but the phase drops are different on all four edges of the elementary plaquette. Note again the high degree of symmetry in the line $\f = \p / 4$: the dynamical stability diagram is completely symmetric, as well as the density modulation strength, also for the striped phase. Apart from the phase distribution, which cannot be symmetric, it is only the ground state energy difference between the checkerboard and striped phases that is different between the two regions $|\f| < \p / 4$ and $|\f| > \p / 4$. In the staggered-vortex region, the checkerboard phase has a lower ground state energy than the striped phase, which can be understood as a consequence of the matching between the sublattice divisions associated with the flux and the density modulation. The flux breaks the symmetry between sublattices $\aa$ and $\bb$ (for details see Ref.~\cite{lk10}), thus introducing a checkerboard pattern, which competes with the striped pattern introduced by the NN interactions.

Also note the shape of the second-order phase boundary, in this case between the homogeneous and striped phases. It shows the same pattern as the boundary between the homogeneous and checkerboard phases (see Fig.~3), and the SF-MI boundary in the absence of NN interactions (see Fig.~1). This shape can be understood from the effect of the flux on the hopping energy in the ground state: it is modified by $\cos \f$ in the uniform superfluid, and $\sin \f$ in the staggered-vortex superfluid, resulting in a minimum at $\f = \p / 4$. Since it is the hopping term in the Hamiltonian that favors the superfluidity in the SF-MI phase transition, and the homogeneity in the SF-CSS and SF-SSS transitions, the reduction in hopping energy makes phases which break the phase coherence or homogeneous density distribution more favorable.

There is one region, close to $\f = \p / 4$, where we do not find any dynamically stable phases (FR - forbidden region). This result is a consequence of our approach, which assumes the existence of a well-defined chemical potential for the whole system, and hence a uniform macroscopic density distribution, since we are not working in a trap. If we drop the assumption of a well-defined chemical potential, we effectively allow the system to separate into parts with different densities. Shifting the density changes the parameter $J / \n U$, which changes the critical values of $V_x / U$ in such a way that any point in the forbidden region can be made to lie within the two closest neighboring stable regions. Hence, if the parameters are tuned to lie in the forbidden region, our calculations predict that the system will phase separate. Note that there are actually two different forbidden regions, since the phases the system will separate into are different for $|\f| < \p / 4$ and $|\f| > \p / 4$.

\subsection{Phase diagram cross section II: $\f = \p / 2$}\label{secfipi2}
Finally, a comment on the two cross sections of constant $\f$ (see Fig.~2). We show the cases where $\f = 0$ and $\f = \p / 2$, {\it i.e.}~the centers of the uniform and the staggered-vortex phases. As mentioned above, the phase diagram has a high degree of symmetry in the line $\f = \p / 4$, the only differences being the phase distribution and the ground state energies of the striped and checkerboard phases. In Fig.~2, we see that the CSS-SSS phase boundary shifts, but nothing else. Intermediate cross sections would also reveal the forbidden region discussed above, and the disappearance of the striped phase near $\f = \p / 4$.

\section{Experimental signatures}\label{secexpsig}

A good starting point for experimental detection of the various phases discussed in this paper is the momentum distribution $n(\vc{k})$, since most of the phases have a unique momentum distribution, as will be discussed below. Experimentally, the momentum distribution is accessible through the technique of time-of-flight measurement \cite{bloch08}, which converts the momentum distribution into a spatial one by suddenly turning off the lattice and allowing the cloud to expand ballistically. $n(\vc{k})$ is given by \cite{lk10}
\begin{align*}
n(\vc{k}) = & \, |w(\vc{k})|^2 \biggl| \sum_\vc{P} e^{i \vc{k} \cdot \vc{P}} \biggr|^2 \sum_{\m, \n = 1}^4 e^{i \vc{k} \cdot (\vc{r}_\n - \vc{r}_\m)} \avg{a^\dag_\vc{r_\m} a_\vc{r_\n}}.
\end{align*}
Here, $\vc{P}$ runs over all plaquettes, i.e. $\vc{P} = 2 a (n_x, n_y)$, with integer $n_{x/y}$, and $\vc{r}_\m$ and $\vc{r}_\n$ run over the four sites of the plaquette. The last factor is the one that distinguishes between the phases, since all configurations are invariant under translation by $\vc{P}$. To calculate it, we follow Ref.~\cite{lk10} and approximate the $N$-particle state on the lattice by a coherent state with on average $N$ particles. Now the calculation becomes very simple, since in this approximation,
\begin{align*}
\avg{a^\dag_\vc{r_\m} a_\vc{r_\n}} \approx \avg{a_\m} \avg{a_\n},
\end{align*}
for which we have explicit results (see sections \ref{secmethod}-\ref{secasymm}). Using the same parameters as in Ref.~\cite{lk10}, we obtain the signatures shown in Figs.~5 and 6.
\begin{figure}
\begin{center}
\includegraphics[width=.2\textwidth]{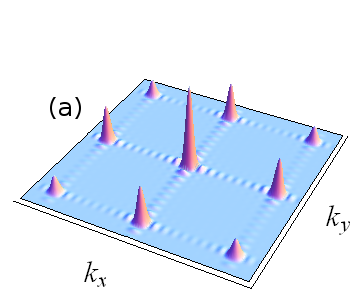}
\includegraphics[width=.2\textwidth]{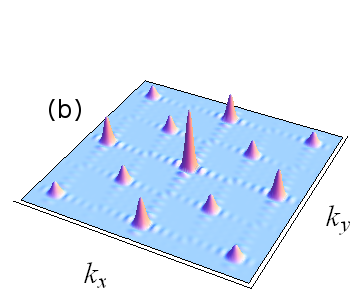}
\includegraphics[width=.2\textwidth]{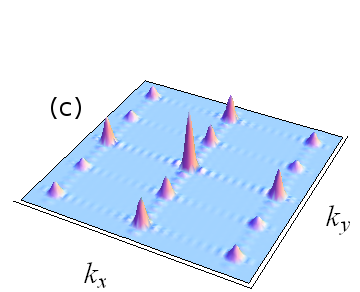}
\end{center}
\caption{The momentum distributions for the (a) SF, (b) CSS, and (c) SSS phases.}
\end{figure}
In Fig.~5, we see that supersolidity manifests itself in the momentum distribution by replacing the peaks from the homogeneous SF by a smaller peak and `satellite peaks' displaced by the characteristic vectors of the density modulation. Fig.~6 shows that the same replacement takes place in the staggered-vortex phases. Unfortunately, this implies that the SVSF and SVCSS phases look exactly the same, since the peaks of the SVSF momentum distribution are displaced from each other by exactly the characteristic vectors of the CSS density modulation.
\begin{figure}
\begin{center}
\includegraphics[width=.2\textwidth]{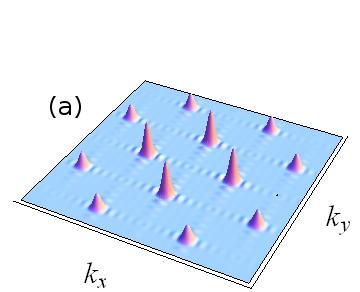}
\\
\includegraphics[width=.2\textwidth]{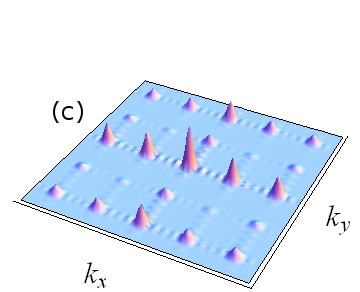}
\includegraphics[width=.2\textwidth]{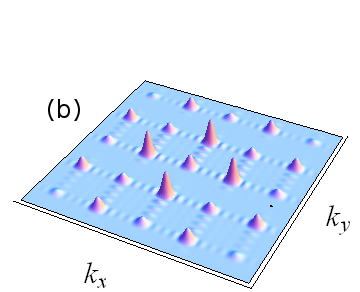}
\end{center}
\caption{The momentum distributions for the (a) SVSF, (b) SVSSS at $\f = 7 \p / 20$, and (c) SVSSS at $\f = \p / 2$ phases. The SVCSS phase has exactly the same signature as the SVSF phase.}
\end{figure}
In Fig.~6, we see another curiosity: the two striped phases, with stripes in the $x$- and $y$-directions, are indistinguishable for $\f = \p / 2$, where the phase drops along the plaquette are equal. For different values of $\f$, the two striped phases are distinguishable. Also, as discussed in sections \ref{secmethod} and \ref{secasymm}, the phase distribution is asymmetric for $\f \notin \{0, \p / 2\}$ in the striped phase. This asymmetry reflects the interplay between NN interactions and applied flux, and is a continuous function of $\f$.

\section{Discussion \& conclusions}\label{secconclude}

In this paper, we have analysed the interplay between NN interactions and a synthetic staggered magnetic field in a system of bosons in a two-dimensional square optical lattice. We have used the Bogolyubov approximation to obtain the theoretical mean-field phase diagram of the system. The equilibrium condition that traditionally gives the value of the chemical potential was replaced by a set of conditions that give the density and phase modulations between the lattice sites, as well as the chemical potential. The excitation spectrum allowed us to determine the dynamical stability of the phases encountered in the system.

Our analysis resulted in a rich phase diagram featuring various superfluid and supersolid phases. Apart from the conventional and staggered-vortex superfluids and the checkerboard and striped supersolids found before, the system turns out to feature phases which combine a staggered-vortex phase configuration and supersolidity. Where the density modulation is invariant under exchange of the lattice vectors, it does not influence the phase configuration; in case of an asymmetric density modulation, the phase drops around the plaquette are also distributed asymmetrically. Lastly, we have identified a forbidden region, where the system cannot form a stable state with the same average density in all areas. In this region of parameter space, our calculations predict that the system will phase separate.

We observe that even a rather crude approximation to the dipolar interaction, where the long tail is cut off beyond the NN range, combined with the staggered flux, leads to a very rich phase diagram. Apart from the expected `superposition' of density modulations and a staggered-vortex pattern, another layer of structure emerges: the phase drops do not have to be distributed homogeneously along the elementary plaquette.

We also see that many of the phase transitions can be driven by tuning either the NN interaction strengths or the flux, which is a consequence of the fact that the flux modifies the hopping energy. Thus, it affects both the vorticity, which is a discrete variable, and the density modulation, which is a continuous variable. This is yet another example of the interesting physics that comes with the possibility of generating an artificial staggered magnetic field in an optical lattice.

Potentially interesting questions that were beyond the scope of this paper include taking into account the effects of the long tail of the dipolar interaction and the effects of finite temperatures on the supersolid phases. Since our work was exploratory in nature, we have not been able to address these problems, but they would certainly be relevant for experimental tests of the predicted phases. Another point left unaddressed here is the full periodicity in $\f$ of the phase diagram. Fig.~1 shows the $2 \p$-periodic nature of the system, while we have only considered values of $\f$ between $0$ and $\p / 2$ in this paper. Note that since the two staggered-vortex phases are identical, the system is symmetric under $\f \to -\f$. Thus, we have studied half of the phase diagram's entire period. The other half is expected to feature phases and phenomena similar to those already discussed here, as can be deduced from the nature of the density and phase distributions found.

\section*{Acknowledgements}
We acknowledge financial support from the Netherlands Organization for Scientific Research (NWO). We thank L.K.~Lim and A.~Hemmerich for stimulating discussions.

\end{document}